\documentclass[lettersize,journal]{IEEEtran}

\usepackage{amsmath, graphicx, amssymb, epsfig, color, cite}
\usepackage{array}
\usepackage{multirow}
\usepackage{enumerate}
\usepackage{booktabs}
\usepackage{verbatim}
\usepackage{optidef}
\usepackage{caption}
\usepackage{subcaption}
\usepackage{algorithm}
\usepackage{algpseudocode}
\usepackage{algcompatible}
\algblockdefx{FORP}{ENDFORP}[1]%
  {\textbf{for }#1 \textbf{do in parallel}}%
  {\textbf{end}}
\algblockdefx{FOR}{ENDFOR}[1]%
  {\textbf{for }#1 \textbf{do}}%
  {\textbf{end}}
\algblockdefx{IF}{ENDIF}[1]%
  {\textbf{if }#1 \textbf{then}}%
  {\textbf{end}}
\algblockdefx{FORALL}{ENDFORALL}[1]%
  {\textbf{for all}#1 \textbf{do}}%
  {\textbf{end}}
  
\algblockdefx{IF}{ENDIF}[1]%
  {\textbf{if }#1 \textbf{then}}%
  {\textbf{end}}
\algblockdefx{WHILE}{ENDWHILE}[1]%
  {\textbf{while }#1 \textbf{do}}
  {\textbf{end}}
  
\algnewcommand\algorithmicprocedure{\textbf{function}}
\algnewcommand\FUNC{\item[\algorithmicprocedure]}%
\algnewcommand\algorithmicendprocedure{\textbf{end function}}
\algnewcommand\ENDFUNC{\item[\algorithmicendprocedure]}%
\algnewcommand\algorithmicforeach{\textbf{for each}}
\algdef{S}[FOR]{FOREACH}[1]{\algorithmicforeach\ #1\ \algorithmicdo}

\usepackage{bm}
\usepackage{float}
\usepackage{makecell}
\usepackage{graphicx}
\usepackage{lipsum}
\usepackage{capt-of}

\newcommand{\argmin}{\operatornamewithlimits{argmin}}
\makeatletter
\newcommand{\vast}{\bBigg@{4.5}}
\newcommand{\Vast}{\bBigg@{7.5}}
\makeatother

\begin{document}
    \title{Online Data Generation for MIMO-OFDM Channel Denoising: Transfer Learning vs. Meta Learning} 
    \author{Sungyoung Ha, Ikbeom Lee, Seunghyeon Jeon, and Yo-Seb Jeon 
    \thanks{Sungyoung Ha, Seunghyeon Jeon, and Yo-Seb Jeon are with the Department of Electrical Engineering, POSTECH, Pohang, Gyeongbuk 37673, South Korea (e-mail: sungyoungha@postech.ac.kr; seunghyeon.jeon@postech.ac.kr; yoseb.jeon@postech.ac.kr).}
    \thanks{Ikbeom Lee is with Samsung Research, Samsung Electronics Co., Ltd., Seoul 06765, South Korea (e-mail: ib.lee@samsung.com).}

    }
 
	\vspace{-2mm}
	
	\maketitle
	\vspace{-12mm}
 
    \begin{abstract}  
        Channel denoising is a practical and effective technique for mitigating channel estimation errors in multiple-input multiple-output orthogonal frequency-division multiplexing (MIMO-OFDM) systems. However, adapting denoising techniques to varying channel conditions typically requires prior knowledge or incurs significant training overhead. To address these challenges, we propose a standard-compatible strategy for generating online training data that enables online adaptive channel denoising. The key idea is to leverage high-quality channel estimates obtained via data-aided channel estimation as practical substitutes for unavailable ground-truth channels. Our data-aided method exploits adjacent detected data symbols within a specific time-frequency neighborhood as virtual reference signals, and we analytically derive the optimal size of this neighborhood to minimize the mean squared error of the resulting estimates. By leveraging the proposed strategy, we devise two channel denoising approaches, one based on transfer learning, which fine-tunes a pre-trained denoising neural network, and the other based on meta learning, which rapidly adapts to new channel environments with minimal updates. Simulation results demonstrate that the proposed methods effectively adapt to dynamic channel conditions and significantly reduce channel estimation errors compared to conventional techniques.
    \end{abstract}


    \section{Introduction}\label{Sec:Intro} 
    Modern wireless communication systems require high data rates, low latency, and robust performance under diverse channel conditions. To meet these demands, multiple-input multiple-output (MIMO) and orthogonal frequency-division multiplexing (OFDM) have been widely adopted. MIMO exploits spatial diversity and multiplexing gain through multiple antennas, while OFDM mitigates inter-symbol interference caused by multipath propagation. Their combination, MIMO-OFDM, serves as the backbone of 5G New Radio (NR) and other advanced communication standards \cite{mimo,ofdm,mimo-ofdm}.
    
    A critical factor of MIMO-OFDM is the acquisition of accurate channel state information at the receiver (CSIR). In practical systems, pilot-aided channel estimation is performed using demodulation reference signals (DM-RSs), which are sparsely embedded in the time-frequency resource grid \cite{dmrs_standards}. Linear estimation techniques such as least squares (LS) and linear minimum mean square error (LMMSE) are applied to the DM-RS positions, followed by interpolation to estimate the full channel \cite{mimo-ofdm ce}. However, due to the limited number of pilots and the presence of noise, these initial estimates are often inaccurate, especially in low signal-to-noise ratio (SNR) or rapidly time-varying environments. Since interpolation relies on these noisy pilot estimates, error propagation is inevitable, leading to performance degradation in data detection \cite{ce_filter1, mismatch}.
    
    To address this issue, channel denoising techniques have been introduced to refine the initial channel estimates and suppress the noise component. Recent deep learning (DL) approaches have demonstrated the ability to learn complex channel characteristics and outperform traditional filters, because it can handle the nonlinearites in the channel effectively. However, most of these methods struggle to adapt to changing channel conditions without retraining, but the true channels are limited. This paper proposes an online adaptive channel denoising framework that eliminates the need for ground-truth channels by generating training data using data-aided channel estimation. We compare two adaptation strategies, transfer learning and meta learning for enabling fast and effective adaptation to dynamic wireless environments using a pre-trained model.

    \subsection{Related Prior Works}
    Data-aided channel estimation has been extensively studied as an effective methods to improve channel estimation accuracy in MIMO-OFDM systems. The fundamental idea is to treat detected data symbols as additional reference signals, thereby increasing the effective number of pilots available for estimation \cite{da_ce, da_ce_rl}. By leveraging this approach, the receiver can achieve improved channel estimates compared to those obtained solely from predefined DM-RSs.

    A major class of data-aided channel estimation techniques is based on iterative estimation. In these methods, the receiver alternates between channel estimation, data detection, and decoding across multiple iterations. The refined data symbols obtained from each iteration are treated as virtual pilots in the next. For instance, \cite{da_ce_iter1} uses soft decisions reconstructed via log-likelihood ratios (LLRs) from the decoder. In \cite{da_ce_iter2}, only a subset of soft decisions is selected based on a mean squared error (MSE)-driven criterion to avoid error propagation. Similarly, \cite{da_ce_iter3} proposes a joint estimation and decoding framework using expectation propagation, reducing both channel and symbol errors iteratively.
    
    While these iterative methods provide notable performance gains, they incur high computational complexity due to repeated decoding and estimation steps. As such, they are less practical in real-time or low-power communication systems.
    To alleviate this burden, non-iterative data-aided estimation has also been explored. These methods refine the channel estimate using detected data from a single pass. For example, \cite{da_ce1} simply augments pilot symbols with reconstructed data symbols. However, undetected errors may introduce bias. To address this, \cite{da_ce2} proposes a semi-data-aided approach that selects only reliable data symbols based on a reinforcement learning-based metric. While this avoids iterative decoding, it still introduces nontrivial complexity in evaluating reliability and requires re-estimating the channel.
    
    An alternative direction is channel denoising, which focuses on suppressing the noise in the initially estimated channel rather than altering the estimation process. Denoising is applied only once after channel estimation, avoiding the need for iterative decoding or detection. Classical methods, such as Wiener filtering \cite{ce_denoising_filter1}, exploit temporal correlations to smooth out estimation noise and have been extended to adaptive filtering for time-varying channels. However, designing optimal filters typically requires prior knowledge of channel or noise statistics. Moreover, mismatches between assumed and actual statistics can significantly degrade performance.
    
    Recent advances in DL have introduced powerful alternatives for channel denoising \cite{dncnn, nlrn, n2n}. Neural networks can learn complex mappings from noisy to clean channel estimates, trained on large datasets without the need for explicit statistical modeling. Motivated by image denoising, many works model the channel frequency response (CFR) as a two dimensional (2D) image over time and frequency. For instance, \cite{ce_denoising2} proposes a two-stage framework combining super-resolution and image restoration. \cite{ce_denoising3} introduces a noise-level estimation module to improve robustness under varying SNRs. The method in \cite{ce_denoising1} embeds learned denoisers into iterative recovery algorithms, while \cite{ce_denoising4} uses deep image priors to boost performance with limited data. Transformer-based approaches have also been proposed \cite{ce_denoising6}, showing scalability and effectiveness in complex settings.
    Despite their promise, DL-based denoising methods face critical challenges. Most notably, they require a large amount of labeled training data that accurately reflects real-world wireless channel conditions. Also, in practical systems, ground-truth CFRs are unavailable, and channel characteristics vary widely with user mobility, location, and time. This domain mismatch between training and deployment environments can severely degrade performance.

    Transfer learning and meta learning have emerged as promising approaches for mitigating the domain mismatch problem and enabling rapid model adaptation. Transfer learning primarily involves fine-tuning a pre-trained model using data from similar domain \cite{transfer}, whereas meta learning targets the identification of a generalized model initialization that can be swiftly adapted with minimal data \cite{meta}. Motivated by these advantages, recent studies have explored the integration of transfer learning and meta learning into wireless communication systems \cite{transfer_2, transfer_3, transfer_4, meta_2, meta_3, meta_4}. However, despite the demonstrated effectiveness of these adaptation techniques in various domains, their potential benefits for enabling online adaptive channel denoising in MIMO-OFDM systems remain unexplored. Additionally, developing a realistic approach to acquire suitable online training data necessary for these techniques continues to be an open challenge.


    \subsection{Our Contributions}
    To bridge this research gap, we propose a standard-compatible strategy for generating online training data in MIMO-OFDM systems. Leveraging this strategy, we develop two online adaptive channel denoising approaches based on transfer learning and meta learning, respectively. Both approaches exploit the generated online dataset to facilitate efficient adaptation of denoising neural networks to online channel conditions. Simulation results confirm that these approaches significantly improve channel estimation accuracy by effectively reducing estimation errors under varying channel environments. The main contributions of this paper are summarized as follows:

    \begin{itemize}
    \item We propose a practical and standard-compatible strategy to generate online training data for channel denoising in MIMO-OFDM systems. A main challenge is that the true CFRs, which serve as labels, are not available at the receiver during online communication. To overcome this, we replace the unavailable ground-truth CFRs with \textit{high-quality} channel estimates obtained via data-aided channel estimation. Specifically, we introduce a low-complexity data-aided method that utilizes adjacent detected data symbols within a designated time-frequency range as virtual DM-RSs. We analytically derive the optimal size of this range to minimize the MSE of the resulting channel estimates. Importantly, our approach requires no additional pilot overhead, ensuring compatibility with existing wireless standards.
    
    \item Based on the proposed data generation strategy, we present a transfer learning approach for online adaptive channel denoising in MIMO-OFDM systems. This approach comprises two phases: pre-training and fine-tuning. During pre-training, a denoising neural network is trained offline using diverse channel characteristics. In the fine-tuning phase, the denoising neural network is adapted using the online data generated by our strategy. This allows the model to quickly adjust to changing channel conditions with minimal computational burden and without introducing signaling overhead.
    
    \item We also present a meta learning approach for online adaptive channel denoising in MIMO-OFDM systems. Unlike transfer learning, this approach enables the denoising neural network to rapidly adapt to new channel environments using only a few gradient updates with a small number of data. The model is meta-trained across a wide variety of channel environment tasks so that it generalizes effectively in out-of-distribution scenarios.
    
    \item We validate the effectiveness of the proposed approaches through extensive simulations. Our results demonstrate that our denoising approaches outperform conventional estimation techniques by significantly mitigating channel errors. Furthermore, the proposed data generation strategy achieves performance comparable to the ideal scenario in which true CFRs are available as training labels.
    \end{itemize}

    This work builds upon our previous study \cite{ICC: Submission}, in which we only introduced a meta learning approach for online adaptive channel denoising in MIMO-OFDM systems. In the present study, we extend this line of work by optimizing the performance of the online training data generation strategy through careful adjustment of the time-frequency range used in data-aided channel estimation. Furthermore, we introduce a novel transfer learning approach, which differs fundamentally from the meta learning approach in terms of its pre-training and fine-tuning procedures. In addition, we enhance the simulation study by providing an in-depth comparative analysis of the transfer and meta learning approaches under various training and testing scenarios.

    \vspace{2mm}
    {\em Notation:} Scalars are denoted by lowercase letters (e.g., $a$), vectors by bold lowercase letters (e.g., $\mathbf{a}$), and matrices by bold uppercase letters (e.g., $\mathbf{A}$).  
    Superscripts $(\cdot)^{-1}$, $(\cdot)^*$, $(\cdot)^{\sf T}$, and $(\cdot)^{\sf H}$ denote the inverse, complex conjugate, transpose, and Hermitian (conjugate transpose), respectively.  
    $\mathbb{E}[\cdot]$ denotes the expectation operator.
    For a scalar $a$, $|a|$ denotes the absolute value.  
    For a vector $\mathbf{a}$, $\|\mathbf{a}\|$ denotes the Euclidean norm, and $(\mathbf{a})_i$ denotes its $i$-th element.  
    For a matrix $\mathbf{A}$, $(\mathbf{A})^{(i,j)}$ denotes the element in the $i$-th row and $j$-th column, and $\|\mathbf{A}\|_{\mathrm{F}}$ denotes the Frobenius norm.  
    $\mathbf{I}_m$ represents the $m \times m$ identity matrix.  
    $\mathcal{CN}(\boldsymbol{\mu}, \mathbf{C})$ denotes the circularly-symmetric complex Gaussian distribution with mean vector $\boldsymbol{\mu}$ and covariance matrix $\mathbf{C}$.

    \section{System Model and Preliminaries}\label{Sec:SystemModel} 
    In this section, we introduce the MIMO-OFDM communication system and discuss the associated challenges in channel estimation.

    \subsection{MIMO-OFDM Communication System}

    Consider a MIMO-OFDM communication system in which a transmitter equipped with \(N_{t}\) antennas communicates with a receiver equipped with \(N_{r}\) antennas over \(K\) consecutive subcarriers. We assume that a resource element (RE), defined as the smallest unit of time-frequency resource, consists of a single subcarrier within one OFDM symbol. Given a power delay profile corresponds to the channel impulse response from the 3GPP standard \cite{cdl}, the CFR for the $(n,k)$-th RE is modeled as
    \begin{align}
        {\bf H}[n,k] = \sum_{l=0}^{L-1} a_{l}[n]e^{-j2\pi k \Delta_{f}\tau_{l}},
    \end{align}
    where $\Delta_{f}$ denotes the subcarrier spacing, and $a_{l}$ and $\tau_{l}$ are the complex coefficient and path delay of the $l$-th path, respectively.
    
    \subsubsection{DM-RS Transmission}
\begin{figure}[t]
        \centering
        \includegraphics[width=4.5cm]{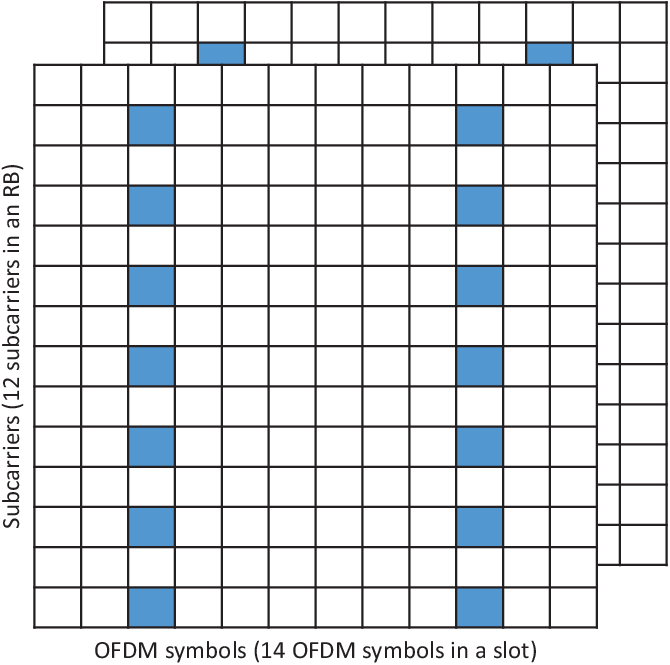}
        \caption{A single-symbol DM-RS employed in 5G NR when configuration type 1 and mapping type A are considered with one additional reference signal occasion.}
        \vspace{-3mm}
        \label{fig:DMRS config.}
    \end{figure}

    To enable channel estimation at the receiver, we assume that some REs are allocated for the transmission of DM-RSs, as in 5G NR. 
    The DM-RSs are known at both the transmitter and receiver, and thus can be utilized for channel estimation. Let $\mathcal{I}_{p} \subset \{(n, k)\}_{\forall n, k}$ be the index set of the REs assigned for DM-RS transmission. Also, let ${\bf x}_p[n, k] = \left[x_{p,1}[n, k], \ldots, x_{p, N_t}[n, k]\right]^{\sf T} \in \mathbb{C}^{N_t}$ be the DM-RS transmitted using the $(n, k)$-th RE for all $(n, k) \in \mathcal{I}_p$. Each DM-RS is assumed to be normalized such that $\mathbb{E}[|x_{p, i}[n, k]|^2] = 1$. Then, for $(n, k) \in \mathcal{I}_p$,  the corresponding signal at the $(n,k)$-th RE is given by  
    \begin{align}
        {\bf y}[n,k] = {\bf H}[n,k]{\bf x}_p[n,k] + {\bf v}[n,k],
    \end{align}
    where ${\bf v}[n,k] \sim \mathcal{CN}\left({\bf 0}_{N_{r}},\sigma^2{\bf I}_{N_{r}}\right)$ is an independent circularly symmetric complex Gaussian noise vector, and $\sigma^2$ is a noise variance.
    Fig.~\ref{fig:DMRS config.} illustrates a specific DM-RS configuration used in this work, based on 5G NR with configuration type 1 and mapping type A, blue squares indicate the DM-RS positions \cite{dmrs_standards}.

    \subsubsection{Data Transmission}
    We assume that the REs not utilized for the DM-RS transmission are assigned for the purpose of data transmission. 
    Let $\mathcal{I}_d \subset \{(n,k)\}_{\forall n,k}$ be the index set of the REs assigned for the data transmission.
    Also, let ${\bf x}[n,k] = \left[x_1[n,k],\ldots, x_{N_t}[n,k]\right]^T \in \mathcal{X}^{N_t}$ be a data symbol vector transmitted using the $(n,k)$-th RE for all $(n,k) \in\mathcal{I}_d$, where $\mathcal{X}$ is a constellation set. Each data signal is assumed to be normalized such that $\mathbb{E}[|x_i[n,k]|^2] = 1$. Then, for $(n,k) \in\mathcal{I}_d$, a received signal at the $(n,k)$-th RE is represented as
    \begin{align}
        {\bf y}[n,k] = {\bf H}[n,k]{\bf x}[n,k] + {\bf v}[n,k].
    \end{align}
    The goal of the receiver is to detect the transmitted data ${\bf x}[n,k]$ from the received signal ${\bf y}[n,k]$ based on the estimated CFR $\hat{\bf H}[n,k]$.

    \begin{figure*}[t]
        \centering
        \includegraphics[width=16cm]{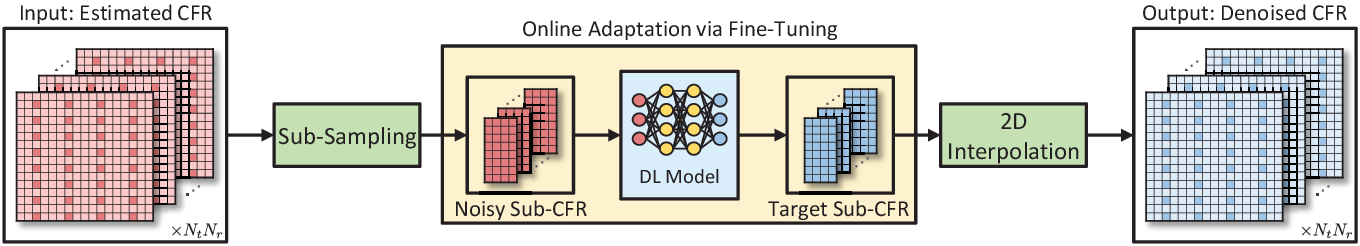}
        \caption{An illustration of DL-based channel denoising approaches applied to sub-sampled CFR estimates.}
        \vspace{-2mm}
        \label{fig: proposed process}
    \end{figure*}

    \subsection{Channel Estimation based on DM-RS}\label{eq:AddModel}
    A typical channel estimation method in MIMO-OFDM systems involves estimating the CFRs at DM-RS positions (i.e., $(n,k) \in \mathcal{I}_p$) using the known DM-RSs $\{{\bf x}_p[n,k]\}_{(n,k)\in\mathcal{I}_p}$, followed by 2D interpolation to obtain CFRs at the data positions (i.e., $(n,k) \in \mathcal{I}_d$). Linear estimation methods such as LS and LMMSE are widely used due to their simplicity and effectiveness in high-SNR conditions. However, in low-SNR scenarios, these methods suffer from considerable degradation. This is because the limited number of DM-RS symbols leads to noisy CFR estimates, and the subsequent interpolation further propagates this noise to non-DM-RS positions. Such inaccuracies negatively affect data detection performance and reduce overall system reliability. 
  
    \subsection{DL-based Channel Denoising}
    In this work, we adopt a DL-based channel denoising framework to mitigate channel estimation errors in MIMO-OFDM systems. The core idea of this framework is to interpret noisy CFR estimates as noisy 2D images and apply a denoising neural network (DnNN), originally designed for image denoising, to recover clean CFR estimates. To manage the computational complexity associated with processing large CFR maps, we apply the DL-based channel denoising only on a sub-sampled set of CFRs, termed {\em sub-CFRs}, instead of the entire CFR map. This sub-sampling strategy significantly reduces computational complexity while preserving denoising performance. Subsequently, a 2D interpolation is applied to reconstruct the complete CFR map. The DL-based channel denoising with sub-sampling is illustrated in Fig.~\ref{fig: proposed process}.

    

    Formally, let ${\bf H}^{(i,j)} = {\bf H}[n_i,k_j]$ denote the true sub-CFR at the $(i,j)$-th time-frequency location, where $n_i$ and $k_j$ represent the sub-sampling points in the time and frequency domains, respectively. The corresponding noisy estimate obtained via DM-RS-based channel estimation is denoted by $\hat{\bf H}^{(i,j)} = \hat{\bf H}[n_i,k_j]$. The DnNN operates on a noisy sub-CFR map comprising $M_t$ consecutive sub-CFRs in the time domain and $M_f$ consecutive sub-CFRs in the frequency domain. The $(r,t)$-th noisy sub-CFR map ending at the $(i,j)$-th sub-sampling index is defined as given in \eqref{eq:DNN_input}.
    \begin{figure*}
    \begin{align}\label{eq:DNN_input}
        \hat{\bf M}_{r,t}^{(i, j)} = \begin{bmatrix}
            \hat{h}_{r,t}^{(i-M_t+1, j-M_f+1)}&  \hat{h}_{r,t}^{(i-M_t+2, j-M_f+1)}& \cdots & \hat{h}_{r,t}^{(i, j-M_f+1)} \\
            \hat{h}_{r,t}^{(i-M_t+1, j-M_f+2)} &  \hat{h}_{r,t}^{(i-M_t+2, j-M_f+2)}& \cdots & \hat{h}_{r,t}^{(i,j-M_f+2)} \\
            \vdots &  \vdots & \ddots & \vdots \\
            \hat{h}_{r,t}^{(i-M_t+1, j)} &  \hat{h}_{r,t}^{(i-M_t+2, j)} & \cdots & \hat{h}_{r,t}^{(i, j)}
        \end{bmatrix} \in \mathbb{R}^{M_f \times M_t}.
    \end{align}
    \hrulefill	
    \end{figure*}   
    \noindent 
    Then, the denoising process utilizing the DnNN is expressed as
    \begin{align}
        \tilde{\bf M}_{r,t}^{(i, j)}  = f_{\rm DnNN}(\hat{\bf M}_{r,t}^{(i, j)} ;\theta),
    \end{align}
    where $\tilde{\bf M}_{r,t}^{(i, j)}$ represents the denoised sub-CFR map, $f_{\rm DnNN}(\cdot;\theta)$ is the neural network function, and $\theta$ denotes the model parameters. The proposed framework is not limited to a specific DnNN architecture and can be integrated with various neural network structures, including convolutional neural networks, residual networks and transformer-based architectures.

    Most existing DL-based channel denoising techniques rely on offline training using large datasets representative of the deployment environment to achieve effective inference. However, accurately anticipating testing environments is challenging due to the inherently dynamic and unpredictable nature of wireless channels. Consequently, these offline-trained models often experience performance degradation under distribution mismatch or out-of-distribution conditions.

    \subsection{Online Adaptive Channel Denoising}   
    To address this critical issue, in this work, we study online adaptive channel denoising for MIMO-OFDM systems. This approach involves real-time adaptation of the DnNN models to the current, potentially unseen, channel conditions. A major challenge in the online adaptive approach is obtaining ideal training samples that accurately represent the mapping between noisy sub-CFR estimates and their true counterparts, as true sub-CFRs are not directly accessible during online communication due to inevitable channel estimation errors.
    
    To overcome this obstacle, in Sec.~\ref{Sec: OnlineGen}, we present a standard-compatible online data generation strategy employing data-aided estimation techniques to construct high-quality labeled sub-CFR pairs in real-time. 
    Then, in Sec.~\ref{Sec: Learning}, we devise two adaptive learning approaches that leverage the proposed online dataset. 
    Finally, in Sec.~\ref{Sec:simulation}, we demonstrate that our approaches effectively overcome the limitations of conventional offline training and mitigate the performance mismatch typically observed between training and deployment environments in practical MIMO-OFDM systems. 

    \section{Proposed Online Data Generation Strategy for Channel Denoising}\label{Sec: OnlineGen}
    In this section, we propose an online data generation strategy that enables a standard-compatible training sample collection for realizing online adaptive channel denoising in MIMO-OFDM systems. 
     
    \begin{figure*}[t]
        \centering
        \includegraphics[width=16cm]{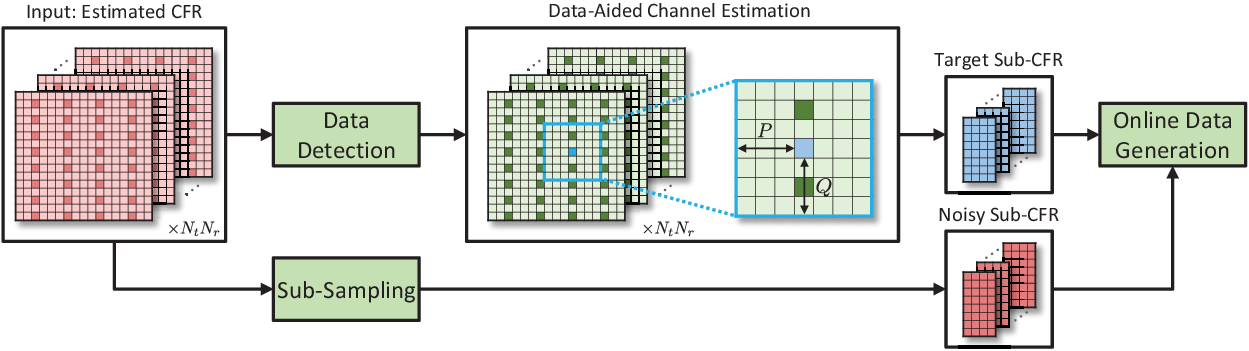}
        \caption{An illustration of the proposed online data generation strategy for channel denoising.}
        \vspace{-3mm}
        \label{fig: DA channel estimation}
    \end{figure*} 
    
    \subsection{Label Generation via Data-aided Channel Estimation}\label{Sec: DA_CE}
    The key idea of the proposed online data generation strategy  is to employ data-aided channel estimation to generate {\it high-quality} sub-CFR estimates, which serve as practical substitutes for the true sub-CFRs in constructing training labels. Data-aided channel estimation has been well known for its effectiveness in reducing channel estimation error by exploiting detected data symbols as virtual DM-RSs \cite{da_ce1,da_ce2}. 

    Inspired by the advantage of data-aided channel estimation, we attain {\it high-quality} sub-CFR estimates by utilizing adjacent detected data symbols as virtual DM-RSs in addition to the original DM-RSs. This process is illustrated in Fig. \ref{fig: DA channel estimation}.
    To be more specific, let \(\hat{\bf x}[n,k]\) be the detected data signal at the \((n,k)\)-th RE for all \((n,k)\in\mathcal{I}_d\) and also let \(P\) and \(Q\) denote the time and frequency ranges used to select adjacent detected data signals.
    Suppose that $P$ and $Q$ are properly determined so that the CFRs within the time-frequency range are highly correlated. The effect and optimization of $P$ and $Q$ will be discussed in Sec. \ref{Sec:Opt_Range}.  
    The CFRs within the time-frequency range centered at the \((i,j)\)-th sub-CFR, denoted by ${\bf H}^{(i,j)} = {\bf H}[n_i,k_j]$, can be approximated as 
    \begin{align}\label{eq: H_approx}
        {\bf H}[n_i - p,k_j - q]  \approx {\bf H}[n_i,k_j], 
    \end{align}    
    for all \(p \in \{-P,\ldots, P\}\) and \(q \in \{-Q, \ldots, Q\}\). 

    Based on the above approximation, the received signal matrix that consists of the received signals within the time-frequency range can be expressed as
    \begin{align}\label{eq:Y_data}
        {\bf Y}_{P,Q}[n_i,k_j] &= \big[{\bf y}[n_i-P,k_j-Q],\cdots,{\bf y}[n_i+P,k_j+Q] \big] \nonumber \\
        &\approx {\bf H}[n_i,k_j]\hat{\bf X}_{P,Q}[n_i,k_j] + {\bf V}_{P,Q}[n_i,k_j],
    \end{align}
    where \(\hat{\bf X}[n_i,k_j]\) consists of the virtual and original DM-RSs in the time-frequency range centered at the \((i,j)\)-th sub-CFR, i.e.,
    \begin{align*}
        \hat{\bf X}_{P,Q}[n_i,k_j] {=} \big[\hat{\bf x}[n_i-P,k_j-Q],\cdots,\hat{\bf x}[n_i+P,k_j+Q]\big], \\
        {\bf V}_{P,Q}[n_i,k_j] {=} \big[{\bf v}[n_i-P,k_j-Q],\cdots,{\bf v}[n_i+P,k_j+Q]\big],
    \end{align*}
    and \(\hat{\bf x}[n,k] = {\bf x}_p[n,k]\) for all \((n,k)\in\mathcal{I}_p\) for notational consistency.
    Then, we determine the data-aided sub-CFR estimate \(\tilde{\bf H}^{(i,j)}\) from the relationship in \eqref{eq:Y_data} by using the conventional channel estimation methods. For example, when employing the LS estimator, the $(i,j)$-th data-aided sub-CFR estimate is computed as 
    \begin{align}\label{eq:DA_CE}
        &\tilde{\bf H}^{(i,j)} \nonumber \\
        &= {\bf Y}_{P,Q}[n_i,k_j]\hat{\bf X}_{P,Q}^{\sf H}[n_i,k_j](\hat{\bf X}_{P,Q}[n_i,k_j]\hat{\bf X}_{P,Q}^{\sf H}[n_i,k_j]  )^{-1}.
    \end{align}    

    Note that the number of virtual DM-RSs used to compute the data-aided sub-CFR estimate \(\tilde{\bf H}^{(i,j)}\) is larger than that used to compute the noisy sub-CFR estimate \(\hat{\bf H}^{(i,j)}\), leading to a significant reduction in the estimation error through a noise averaging effect.
    Therefore, the data-aided sub-CFR estimates can serve as a practical substitute for the true sub-CFRs, facilitating the generation of an online training dataset for optimizing the channel denoising methods.

    \subsection{Optimization for Label Generation}\label{Sec:Opt_Range}
    The effectiveness of the label generation heavily depends on the size of the time-frequency range, denoted by $(P,Q)$, which is used to select adjacent detected data symbols for data-aided channel estimation. When this size increases, the number of virtual DM-RSs for data-aided channel estimation also increases, which may lead to a reduction in channel estimation error. However, if the size becomes too large, the correlations among the CFRs within the time-frequency range decrease, resulting in performance degradation of the data-aided channel estimation. Considering this trade-off, it is essential to optimize the size $(P,Q)$ of the time-frequency range to maximize the quality of the data-aided sub-CFR estimate.

    Motivated by this, we formulate the optimization problem to determine the best size $(P,Q)$ of the time-frequency range that reduces the data-aided channel estimation error. 
    First, we analyze the data-aided channel estimation error as a function of $P$ and $Q$.
    Suppose that the elements of the CFR follow zero mean and variance $\sigma_h^2$.  
    Then a correlation coefficient between the $(n+p,k+q)$-th CFR and $(n,k)$-th CFR is computed as
    \begin{align}
        \epsilon_{p,q} = \frac{\mathbb{E}\big[h_{i,j}[n+p,k+q]h_{i,j}^*[n,k]\big]}{\sigma_h^2}.
    \end{align}
    We assume that the above correlation coefficient does not change across the elements of the CFR. 
    The correlation coefficient $\epsilon_{p,q}$ can be roughly estimated using explicit techniques which is affected from the Doppler effect and multipath propagation \cite{tf_correlation}. Note that high accuracy is not required, as the coefficient is used solely to guide the optimization of $(P,Q)$.
    Utilizing this coefficient, we express the $(n+p,k+q)$-th CFR as a linear function of the $(n,k)$-th CFR as follows:
    \begin{align}
        {\bf H}[n+p,k+q] = \epsilon_{p,q} {\bf H}[n,k] +  {\bf Z}_{p,q}[n,k],
    \end{align}    
    where ${\bf Z}_{p,q}[n,k]$ is a modeling error term which is uncorrelated with ${\bf H}[n,k]$. 
    Let $r_{n,k} = \mathbb{P}({\bf x}[n,k]=\hat{\bf x}[n,k]|{\bf y}[n,k])$ be the a posteriori probability (APP) of the detected data signal $\hat{\bf x}[n,k]$.
    Note that the APP can be obtained can be obtained as part of the data detection process or explicitly computed using the log-likelihood ratios after data detection \cite{soft_app}.
    Since the APP can be interpreted as the reliability of the detected signal at the $(n,k)$-th RE, we approximate the detected data symbol $\hat{x}[n,k]$ at the $(n,k)$-th RE as  
    \begin{align}
        {\bf x}[n,k]  
        &= r_{n,k}\hat{\bf x}[n,k] + {\bf e}[n,k],
    \end{align}
    where ${\bf e}[n,k]$ is a detection error term. In our optimization, we assume that the detection error term is zero mean and uncorrelated with the CFR.

    Under the above model, the received signal at the $(n+p,k+q)$-th RE can be written as
    \begin{align}\label{eq:y_np_kq}
        &\mathbf{y}[n+p, k+q] \nonumber \\
        &= \mathbf{H}[n+p, k+q]\, \mathbf{x}[n+p, k+q] +\mathbf{v}[n+p, k+q]   \nonumber  \\
        &\approx \epsilon_{p,q}\, r_{n+p,q+k}\, \mathbf{H}[n,k]\, \hat{\bf x}[n+p, k+q] +  \tilde{\mathbf{v}}[n+p, k+q],
    \end{align}
    where \(\tilde{\mathbf{v}}[n+p, k+q]\) denotes an effective noise signal defined 
    \begin{align}\label{eq:v_tilde}
    \tilde{\bf v}[n{+}p,k{+}q] = &~{\bf v}[n{+}p,k{+}q] + \epsilon_{p,q}{\bf H}[n,k] {\bf e}[n{+}p,k{+}q] \nonumber \\
    &+r_{n+p,k+q}{\bf Z}_{p,q}[n,k]\hat{\bf x}[n{+}p,k{+}q]  \nonumber \\
    &+ {\bf Z}_{p,q}[n,k] {\bf e}[n{+}p,k{+}q].
    \end{align}
    Note that the effective noise signal includes the effects of the modeling error, detection error, and noise signal. 

    From \eqref{eq:y_np_kq}, the received signal matrix $\mathbf{Y}_{P,Q}[n,k]$ in \eqref{eq:Y_data} which collects the received signals within the time-frequency range centered at the \((n,k)\)-th RE is rewritten as 
    \begin{align}\label{eq:y_PQ}
        {\bf Y}_{P,Q}[n,k] = {\bf H}[n,k]  \hat{\mathbf{X}}_{P,Q}[n,k]  \mathbf{D}_{P,Q}[n,k]  + \tilde{\mathbf{V}}_{P,Q}[n,k],
    \end{align}
    where 
    \begin{align*}
        \tilde{\mathbf{V}}_{P,Q}[n,k]  &= \big[\tilde{\mathbf{v}}[n-P, k-Q], \ldots, \tilde{\mathbf{v}}[n+P, k+Q]\big], \\
        \mathbf{D}_{P,Q}[n,k]  &= \mathrm{diag}(\epsilon_{-P, -Q}r_{n-P, k-Q}, \ldots, \epsilon_{P, Q}r_{n+P, k+Q}),
    \end{align*}  
    and \(\mathrm{diag} (a_1,\ldots,a_I)\) denotes a diagonal matrix whose $i$-th diagonal element is $a_i$.
    To make our optimization mathematically tractable,
    Since the diagonal elements of $\mathbf{D}_{P,Q}[n,k]$ have similar values for a proper range, we further approximate the received signal matrix in \eqref{eq:y_PQ} as
    \begin{align}\label{eq:y_PQ_approx}
        {\bf Y}_{P,Q}[n,k] \approx \xi_{P,Q} {\bf H}[n,k]  \hat{\mathbf{X}}_{P,Q}[n,k]   + \tilde{\mathbf{V}}_{P,Q}[n,k],
    \end{align}
    where
    \begin{align}\label{eq:xi_PQ}
        \xi_{P,Q} = \frac{1}{(2P+1)(2Q+1)} \sum_{p=-P}^{P}\sum_{q=-Q}^{Q} \epsilon_{p, q}  r_{n+p, k+q}.
    \end{align}
   
    In our optimization, we consider the LS-based data-aided channel estimation in \eqref{eq:DA_CE}. Then, the data-aided CFR estimate is given by  
    \begin{align}\label{eq:hat_H_approx}
        \tilde{\mathbf{H}}[n,k] 
        &= {\bf Y}_{P,Q}[n,k]  \hat{\mathbf{X}}_{P,Q}^{\sf H}[n,k]  \big( \hat{\mathbf{X}}_{P,Q}[n,k]  \hat{\mathbf{X}}_{P,Q}^{\sf H}[n,k] \big)^{-1}  \nonumber \\
        &\overset{(a)}\approx \tilde{\mathbf{V}}_{P,Q}[n,k] \hat{\mathbf{X}}_{P,Q}^{\sf H}[n,k]  \big( \hat{\mathbf{X}}_{P,Q}[n,k]  \hat{\mathbf{X}}_{P,Q}^{\sf H}[n,k]  \big)^{-1} \nonumber \\ 
        &~~~+\xi_{P,Q} \mathbf{H}[n,k],
    \end{align}
    where $(a)$ holds from \eqref{eq:y_PQ_approx}. 
    Note that the CFR $\mathbf{H}[n,k]$ and the effective noise signal $\tilde{\bf v}[n+p,k+q]$ are uncorrelated, while the mean of $\tilde{\bf v}[n+p,k+q]$ is assumed to be zero mean. 
    Therefore, the MSE between the true CFR and data-aided CFR estimate is expressed as given in \eqref{eq:MSE}.
    \begin{figure*}[ht]
        \begin{align}\label{eq:MSE}
            \mathbb{E} \big[ \| \mathbf{H}[n,k] - \tilde{\mathbf{H}}[n,k] \|_{\mathrm{F}}^2 \big] 
            &\approx  \left| 1 -\xi_{P,Q}  \right|^2 \mathbb{E} \left[ \| \mathbf{H}[n,k] \|_{\mathrm{F}}^2\right] +  \mathbb{E} \big[ \|\tilde{\bf V}_{P,Q}[n,k] \hat{\mathbf{X}}_{P,Q}^{\sf H}[n,k]  \big( \hat{\mathbf{X}}_{P,Q}[n,k]  \hat{\mathbf{X}}_{P,Q}^{\sf H}[n,k]  \big)^{-1}  \|_{\mathrm{F}}^2\big] \nonumber \\
            &\overset{(a)}{\geq}  \left| 1 -\xi_{P,Q}  \right|^2  \mathbb{E} \left[ \| \mathbf{H}[n,k] \|_{\mathrm{F}}^2\right] + \mathbb{E} \big[ \|{\bf V}_{P,Q}[n,k] \hat{\mathbf{X}}_{P,Q}^{\sf H}[n,k]  \big( \hat{\mathbf{X}}_{P,Q}[n,k]  \hat{\mathbf{X}}_{P,Q}^{\sf H}[n,k]  \big)^{-1}  \|_{\mathrm{F}}^2\big] \nonumber \\
            &= \left| 1 -\xi_{P,Q}  \right|^2 N_r N_t \sigma_h^2 +  N_r\sigma^2{\rm Tr}\big( \big( \hat{\mathbf{X}}_{P,Q}[n,k]  \hat{\mathbf{X}}_{P,Q}^{\sf H}[n,k]  \big)^{-1} \big).
        \end{align}
    \hrulefill
    \end{figure*}
    In this expression, the inequality in $(a)$ holds because all the terms in \eqref{eq:v_tilde} are uncorrelated and zero mean. 

    Based on the MSE analysis in \eqref{eq:MSE}, we determine the optimal size of the time-frequency range to minimize the MSE of the data-aided CFR estimate by solving the following problem:
    \begin{align}\label{eq:opt_size}
        (P^\star, Q^\star)
        &= \underset{P, Q}{\argmin}~  \left| 1 -\xi_{P,Q}  \right|^2   \frac{N_t \sigma_h^2 }{\sigma^2} \nonumber \\
        &~~~~~~~~~~~+ {\rm Tr}\big( \big( \hat{\mathbf{X}}_{P,Q}[n,k]  \hat{\mathbf{X}}_{P,Q}^{\sf H}[n,k]  \big)^{-1} \big).
    \end{align}
    Although this problem depends on the position $(n,k)$ of the target CFR, the optimal size $(P^\star,Q^\star)$ may not vary significantly across different positions. This is because the virtual DM-RS matrix $\hat{\mathbf{X}}_{P,Q}[n,k]$ consists of detected data symbols, which are random and independent of $(n,k)$, and true DM-RSs, which have the same structure across all the entire time-frequency domain. Therefore, we solve the above problem {\em only once} for a representative position (e.g., the first sub-CFR position $(n_1,k_1)$) and apply the resulting size $(P^\star,Q^\star)$ consistently during the data-aided channel estimation process. 

    Our objective function effectively captures the impact of the size $(P,Q)$ of the time-frequency range on the performance of data-aided channel estimation. The first term in the objective function increases with $P$ and $Q$ because the correlation $\epsilon_{p,q}$ among CFRs decreases as $p$ and $q$ increase, leading to a reduction in the value of $\xi_{P,Q}$ defined in \eqref{eq:xi_PQ}. This term, therefore, reflects how the time-frequency correlations of the CFRs influence the performance of data-aided channel estimation. Meanwhile, the second term in the objective function decreases with $P$ and $Q$ because the eigenvalues of $\hat{\mathbf{X}}_{P,Q}[n,k]  \hat{\mathbf{X}}_{P,Q}^{\sf H}[n,k]$ increase with the range size. This term captures the effect of increasing the number of virtual DM-RSs on the data-aided channel estimation performance. The relative importance of these two terms is determined by the term $ \frac{N_t \sigma_h^2 }{\sigma^2}$ which governs the system’s SNR. For instance, when the SNR is low, increasing the number of virtual DM-RSs becomes more critical to compensate for higher channel estimation errors. Taken together, these insights demonstrate that our optimization strategy effectively determines the optimal size of the time-frequency range by balancing the trade-off between maintaining high correlation among CFRs and increasing the number of virtual DM-RSs.

    \begin{figure*}[t]
    \centering
    \begin{subfigure}[b]{0.24\textwidth}
        \centering
        \includegraphics[width=\textwidth]{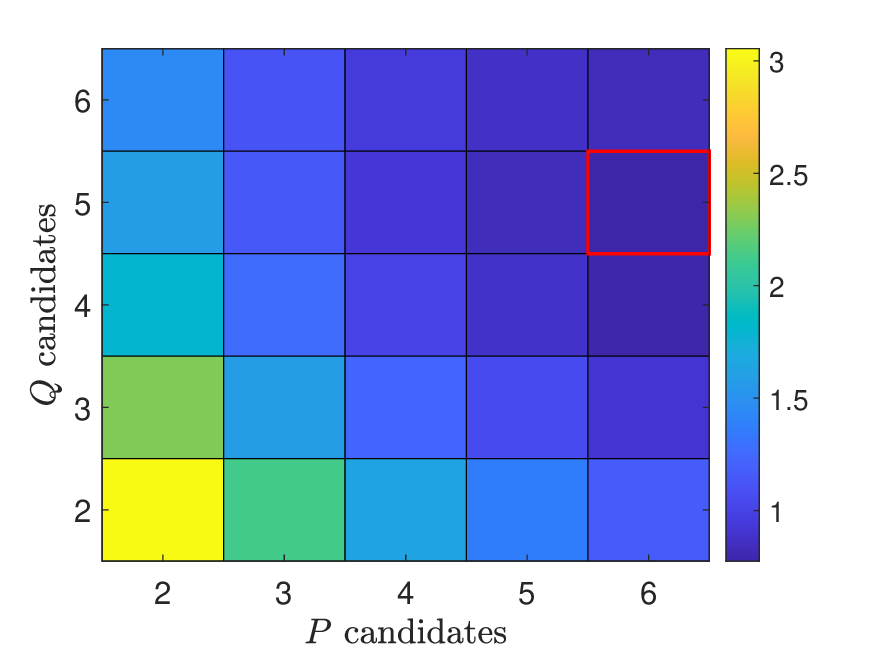}
        \caption{$\mathsf{SNR}=0\mathrm{dB}$ at $v=60\rm{km/h}$}
        \label{fig: 1}
    \end{subfigure}
    \begin{subfigure}[b]{0.24\textwidth}
        \centering
        \includegraphics[width=\textwidth]{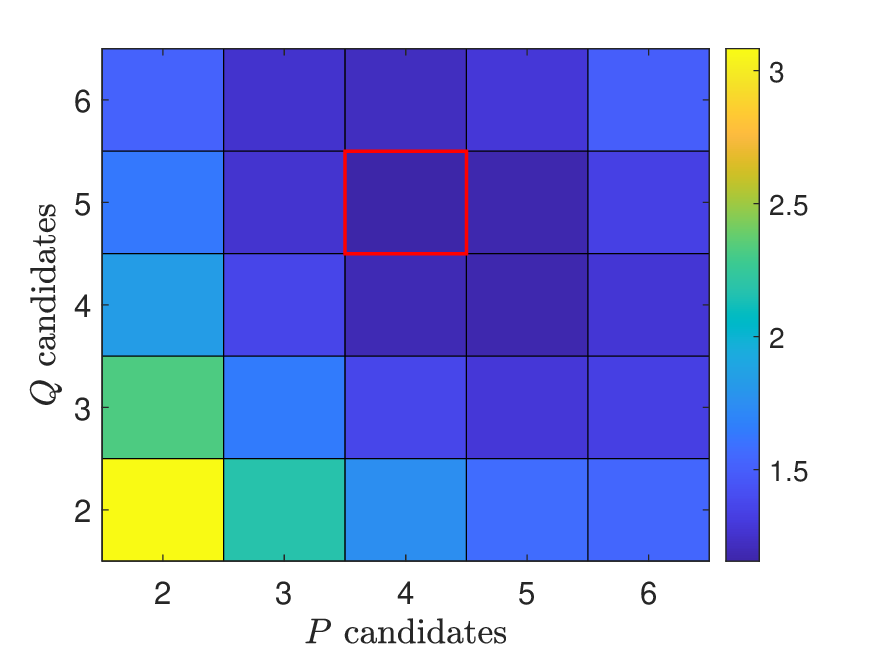}
        \caption{$\mathsf{SNR}=0\mathrm{dB}$ at $v=120\rm{km/h}$}
        \label{fig: 2}
    \end{subfigure}
    \begin{subfigure}[b]{0.24\textwidth}
        \centering
        \includegraphics[width=\textwidth]{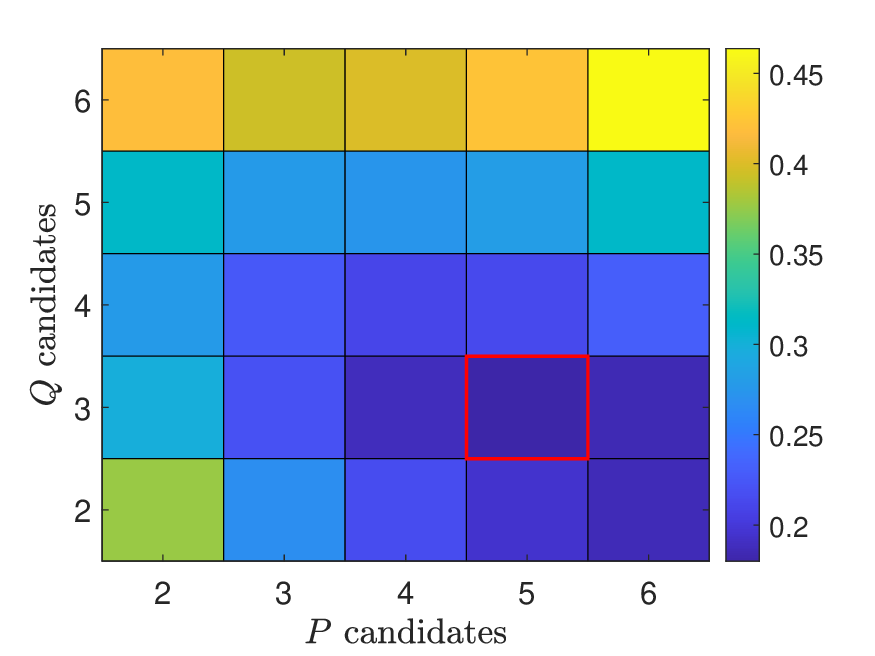}
        \caption{$\mathsf{SNR}=9\mathrm{dB}$ at $v=60\rm{km/h}$}
        \label{fig: 3}
    \end{subfigure}
    \begin{subfigure}[b]{0.24\textwidth}
        \centering
        \includegraphics[width=\textwidth]{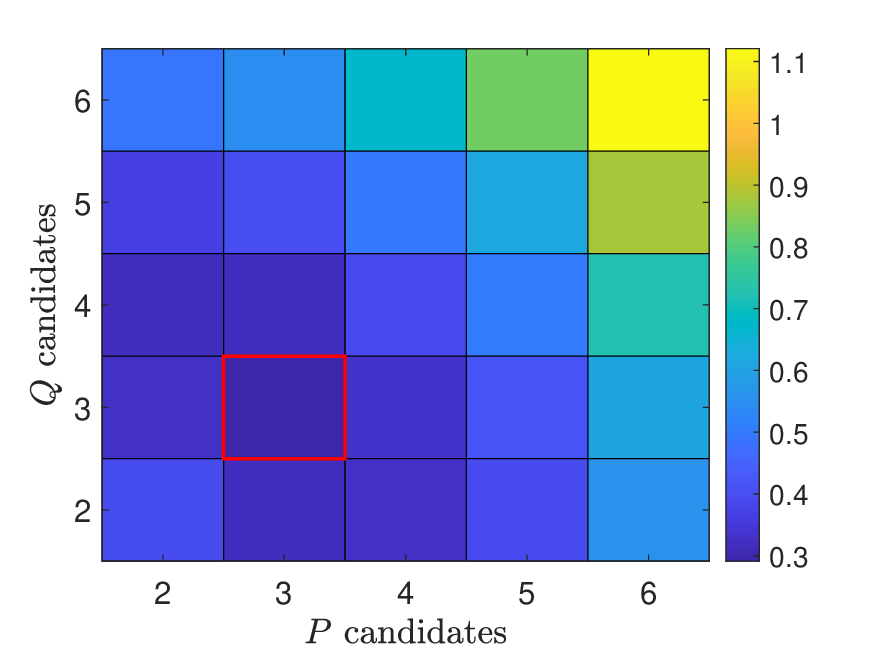}
        \caption{$\mathsf{SNR}=9\mathrm{dB}$ at $v=120\rm{km/h}$}
        \label{fig: 4}
    \end{subfigure}
    
    \caption{MSE values computed from \eqref{eq:MSE} for various window sizes $(P, Q)$ in a CDL-B channel (delay spread = $300{\rm ns}$).}
    \label{fig: MSE_PQ_comparison}
    \vspace{-3mm}
    \end{figure*}

    {\bf Numerical Example (MSE with Various $(P,Q)$):} In Fig. \ref{fig: MSE_PQ_comparison}, we visualize the MSE values computed from \eqref{eq:MSE} for various window sizes $(P, Q)$ in a cluster delay line (CDL) B channel model in \cite{cdl} (delay spread = $300{\rm ns}$). 
    In this simulation, we apply a 2D linear interpolation technique to reconstruct the full CFR map from the denoised sub-CFR map.
    For both DM-RS-based channel estimation  and data-aided channel estimation, we use the LS method, while the LMMSE method is adopted for data detection.
    The results are presented for different SNR and mobility conditions to examine how the optimal window size varies with channel characteristics.
    Figs. \ref{fig: MSE_PQ_comparison} (a) and (b) correspond to the low-SNR scenario ($\mathsf{SNR}=0\rm{dB}$) with the velocities of $60\rm{km/h}$ and $120\rm{km/h}$, respectively. In these cases, the optimal $(P^{\star},Q^{\star})$ values are relatively large, suggesting that a wider time-frequency region should be aggregated to exploit more virtual DM-RSs when the channel observations are highly noisy. 
    By contrast, under the high-SNR scenario ($\mathsf{SNR}=9\rm{dB}$) shown in Figs. \ref{fig: MSE_PQ_comparison} (c) and (d), the optimal window sizes become smaller. This indicates that when the effect of the noise is not severe, the use of a small number of highly-correlated DM-RSs suffices to obtain high-quality channel estimates. A consistent trend is also observed with respect to the mobility because the optimal $P^{\star}$ (time domain extent) tends to decrease as a velocity $v$ increases. This reflects the fact that time correlation decays faster in high-mobility environments, and including too many temporally distant signals can introduce outdated or misleading information into the data-aided channel estimation.

      \begin{figure*}[t]
        \centering
        \includegraphics[width=12cm]{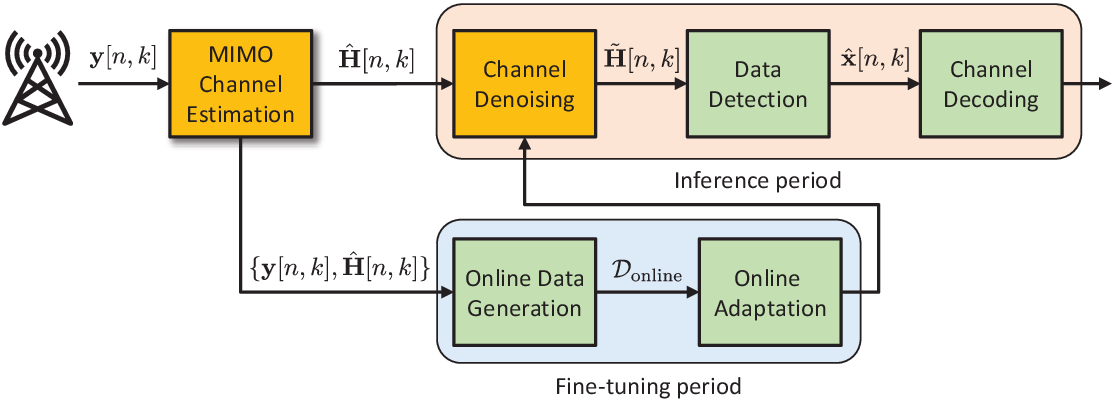}
        \caption{An illustration of the proposed receiver architecture, where the denoising method is adapted using an online training dataset and subsequently used for inference.}
        \vspace{-3mm}
        \label{fig: proposed receiver}
    \end{figure*}

    \subsection{Online Dataset Generation}
    Leveraging the label generation process described above, we generate online training samples, each describing the relationship between the noisy sub-CFR estimate and the corresponding data-aided sub-CFR estimate. To this end, we designate the first \(N_{\text{train}}\) consecutive time slots as a fine-tuning period, during which training samples are generated in real-time. Let \(N_{\text{total}}\) denote the total number of time slots. The remaining \(N_{\text{total}} - N_{\text{train}}\) time slots are defined as an {\it inference} period, during which online data generation is deactivated. We assume that the fine-tuning period is shorter than the inference period, i.e., \(N_{\text{train}} < N_{\text{total}} - N_{\text{train}}\). The proposed receiver architecture is illustrated in Fig. \ref{fig: proposed receiver}.

    
    After the computation, the label (i.e., true sub-CFR) for the \((i,j)\)-th noisy sub-CFR estimate is set as the data-aided CFR estimate \(\tilde{\bf H}^{(i,j)}\) computed for the same sub-sampling position. 
    The obtained samples are arranged into localized 2D windowes of size $M_t \times M_f$ across the time and frequency domains:
    \begin{align}\label{eq:onlineset_DL}
        \mathcal{D}_{\rm online} = \big\{ \big(\hat{\bf M}_{r,t}^{(i, j)}, &\tilde{\bf M}_{r,t}^{(i, j)}\big) \big| r\leq N_{r},t\leq N_{t}, \nonumber \\
        &~~~ n_{i}\leq N_{\rm train}, k_j \leq K \big\},
    \end{align}
    where $\tilde{\bf M}_{r,t}^{(i,j)}$ denotes a data-aided target sub-CFR map.
    Since the dataset is generated during online communication, it inherently reflects the time-varying, user-specific characteristics of the current channel.

    A prominent feature of our strategy is that it does not require additional training overhead beyond the conventional DM-RSs dedicated to channel estimation, thereby ensuring compatibility with modern wireless communication standards. Although this approach introduces extra computational overhead for computing the data-aided estimates, this cost arises exclusively during the fine tuning period. Given that the inference period is significantly longer than the training period, the overall complexity remains manageable. Furthermore, conventional data transmission continues concurrently during the training period, ensuring no loss in spectral efficiency due to the proposed training process.

    \section{Proposed Online Adaptive Channel Denoising Approaches}\label{Sec: Learning}
    By leveraging the online data generation strategy proposed in Sec.~\ref{Sec: OnlineGen}, we propose two online adaptive channel denoising approaches for MIMO-OFDM systems based on transfer learning and meta learning, respectively. These approaches effectively overcome the limitations of conventional offline training and mitigate the performance mismatch typically observed between training and deployment environments in practical MIMO-OFDM systems.

    \subsection{Proposed Transfer Learning Approach}
    In this approach, we employ a transfer learning strategy to mitigate the performance degradation of the DnNN in out-of-distribution scenarios. This strategy enables the DnNN to leverage knowledge acquired from a large offline dataset and subsequently fine-tune its parameters using a small amount of online data reflective of the current wireless channel environment. The proposed transfer learning approach comprises two phases: pre-training and fine-tuning. Each phase is detailed below. 
    \begin{itemize}
        \item \textbf{Pre-Training Phase:}
        In this phase, the DnNN is initialized with random weights and trained on a large offline dataset, denoted as 
        \begin{align}\label{eq: off_data}
            \mathcal{D}_{\rm offline} = \{(\hat{\bf M}_u, {\bf M}_u)\}_{u=1}^{N_{\rm off}},
        \end{align}
         consisting of $N_{\rm off}$ pairs of noisy and true sub-CFR maps. These samples are generated from simulated channel realizations covering a wide range of SNR levels, Doppler frequencies, and delay spreads to mimic realistic MIMO-OFDM propagation scenarios. The pre-training process allows the DnNN to learn general denoising features, such as noise patterns and channel correlation structures across both time and frequency, providing a strong initialization for subsequent adaptation.
    
        ~~~The model is pre-trained by minimizing the empirical loss over the offline dataset:
        \begin{align}\label{eq:DL_off_loss}
            \mathcal{L}_{\rm offline}(\theta) = \frac{1}{N_{\rm off}} \sum_{(\hat{\bf M}_u, {\bf M}_u)\in \mathcal{D}_{\rm offline}} \ell(\hat{\bf M}_u, {\bf M}_u; \theta),
        \end{align}
        where $\ell(\cdot)$ is an MSE loss function that compares the noisy sub-CFR map to the true sub-CFR map.
    
        \item \textbf{Fine-Tuning Phase:}
        Suppose that the pre-trained DnNN is deployed. During this phase, the receiver generates online dataset $\mathcal{D}_{\rm online}$ by leveraging our strategy described in Sec.~\ref{Sec: OnlineGen}. Then, the pre-trained DnNN parameters are fine-tuned using  $\mathcal{D}_{\rm online}$. 
        This adaptation is performed by updating the full network parameters $\theta$ using a small number of gradient steps:
        \begin{align}\label{eq: fine tuning grad}
            \theta \leftarrow \theta - \gamma \nabla_\theta \mathcal{L}(f_\theta),
        \end{align}
        where $\mathcal{L}(f_\theta)$ denotes the loss function computed on the online dataset $\mathcal{D}_{\rm online}$, $\nabla_\theta \mathcal{L}(\cdot)$ is the gradient of the loss with respect to the network parameters, and $\gamma$ is the learning rate used during fine-tuning.
        To mitigate overfitting due to the limited size of the online dataset and the risk of distribution mismatch, we adopt a deep fine-tuning strategy, in which all layers of the pre-trained DnNN are updated with a reduced learning rate and small update steps.
        This strategy is known to preserve generalization learned during pre-training while allowing for sufficient flexibility to adapt to new environments \cite{fine_tuning_1}.
        
    
    
    \end{itemize}

    \subsection{Proposed Meta Learning Approach}
    In this approach, we employ a model-agnostic meta learning (MAML) strategy in \cite{maml} to enhance the generalization capability of the DnNN. MAML aims to learn a set of initial model parameters that can be quickly adapted to unseen tasks using only a few gradient updates with small number of samples. 
    Based on the MAML principle, we construct a meta-regression task using the offline dataset defined in \eqref{eq: off_data}. Each task is divided into a support set and a query set as follows
    \begin{align}
        \mathcal{D}_{{\rm sup}} &= \{(\hat{\bf M}_u, {\bf M}_u)\}_{u=1}^{N_{\rm sup}}, \\
        \mathcal{D}_{{\rm que}} &= \{(\hat{\bf M}_u, {\bf M}_u)\}_{u=1}^{N_{\rm que}},
    \end{align}
    where \( N_{\rm sup} \) and \( N_{\rm que} \) denote the number of support and query samples, respectively, used for each task during the meta-training phase. The support set $\mathcal{D}_{{\rm sup}}$ is used for task-specific fine-tuning for the task (inner update), while the query set $\mathcal{D}_{{\rm que}}$ is used to evaluate the model's generalization performance after adaptation (outer update). Then pass the optimized $\theta$ to the next task. In the context of channel denoising, each task corresponds to a distinct channel condition that may vary in terms of SNR, Doppler spread, angular scattering, or delay profile. By training across such diverse conditions, the model learns denoising features that are transferable across different channel environments.
    
    Based on the MAML, the meta learning-based channel denoising process consists of two main phases: meta-training and meta-testing. Each phase is detailed below. 
    \begin{itemize}
        \item \textbf{Meta-Training Phase:} 
        In this phase, the DnNN parameters $\theta$ are optimized over a distribution of tasks $\{\mathcal{T}_i\}$. For each task $\mathcal{T}_i$, the model is first adapted using the support set via a gradient-based inner update:
        \begin{align}\label{eq:localupdate}
            \theta_{i}^{'} = \theta - \alpha \nabla_{\theta} \mathcal{L}_{\mathcal{T}_i}(f_{\theta}),
        \end{align}
        where $\alpha$ is the inner loop learning rate and $\mathcal{L}_{\mathcal{T}_i}$ is the loss between noisy and clean sub-CFR maps for task $\mathcal{T}_i$.
    
        ~~~The fine-tuned parameters $\theta_i^{'}$ are then evaluated on the query set, and the resulting losses are aggregated to update the global parameters:
        \begin{align}\label{eq:globalupdate}
            \theta \leftarrow \theta - \beta \nabla_{\theta} \sum_i \mathcal{L}_{\mathcal{T}_i}(f_{\theta_{i}^{'}}),
        \end{align}
        where $\beta$ is the outer loop learning rate and $\theta$ is updated using the adaptive moment estimation (ADAM) optimizer. Note that the gradients are computed with respect to the $\theta$ which means that the initial point will be positioned as a good position for fine-tuning. In our implementation, we adopt the MSE as the loss function for both the inner and outer updates, measuring the error between the predicted and target sub-CFR maps.
    
        ~~~This bi-level optimization process enables the model to learn initialization $\theta$ that capture shared representations across diverse channel conditions, while the task-specific parameters $\theta_i^{'}$ capture local variations. As a result, the model becomes capable of quickly adapting to new, unseen channel environments with minimal computation. This meta-training is described in {Algorithm~\ref{alg:MAML}}.
        
        \item \textbf{Fine-Tuning Phase:} 
        During online operation, the BS encounters new tasks (e.g., new environments). In such cases, the online dataset $\mathcal{D}_{\rm online}$ is generated according to the strategy described in Sec. \ref{Sec: OnlineGen} under the current environment. The DnNN is then adapted from the meta-learned initialization $\theta$ via a small number of gradient updates, as defined in \eqref{eq: fine tuning grad}. However, meta learning enables more efficient adaptation by reducing the fine-tuning overhead. This allows the model to rapidly specialize to the current channel conditions using only a limited number of online samples, demonstrating the effectiveness of meta-learning in dynamic wireless environments.
    
    \end{itemize}

    \begin{algorithm}[t]
        \caption{Proposed MAML-based Meta-Training Process}\label{alg:MAML}
        \small
        \begin{algorithmic}[1]
            \REQUIRE $\{\mathcal{D}_u\}$, $\alpha, \beta$
            \ENSURE $\theta$
            
            \State Randomly initialize the DnNN parameters $\theta$.
            \WHILE{not done}
                \State Sample batch of tasks $\mathcal{T}_{i}$.
                \FORALL{ $\mathcal{T}_i$}
                    \State Evaluate $\nabla_{\theta}\mathcal{L}_{\mathcal{T}_i}(f_{\theta})$ using $\mathcal{D}_{\rm sup}$ and $\mathcal{L}_{\mathcal{T}_i}$.
                    \State Compute adapted parameters $\theta_{i}^{'}$ by \eqref{eq:localupdate}. 
                \ENDFORALL
                \State Update $\theta$ by \eqref{eq:globalupdate} with $\mathcal{D}_{\rm que}$ and $\mathcal{L}_{\mathcal{T}_i}$.
            \ENDWHILE
        \end{algorithmic}
    \end{algorithm}

    {\em Remark (Comparison Between Transfer and Meta Learning):}
    It should be noted that although both transfer learning and meta learning involve a two-phase training process, the key distinction lies in that meta learning explicitly optimizes the model for rapid adaptation across diverse channel tasks, whereas transfer learning aims to fine-tune a generally pre-trained model without a task-specific structure.

    \subsection{Online Adaptive Channel Denoising Process}
    The online adaptive channel denoising procedure when employing the proposed approaches is summarized in {Algorithm~\ref{alg:Procedure}}. This procedure consists of two distinct phases: \textit{fine-tuning} and \textit{inference}. We assume the DnNN parameters are pre-trained according to the methods described in the proposed transfer learning or meta learning approaches. 

    During the fine-tuning phase, corresponding to the first $N_{\rm train}$ OFDM symbols in online communications, the receiver constructs an online training dataset $\mathcal{D}_{\rm online}$. The DnNN parameters are then fine-tuned based on this dataset, using either the transfer learning or meta learning procedure. This allows the pre-trained DnNN to dynamically adapt to current channel conditions, ensuring robust denoising performance in practical MIMO-OFDM systems. Importantly, conventional data transmission and reception proceed concurrently during this period, thus preserving the spectral efficiency of the system.

    In the subsequent inference period, covering the remaining $N_{\rm total}-N_{\rm train}$ OFDM symbols, the fine-tuned DnNN transforms each noisy sub-CFR map $\hat{\mathbf{M}}_{r,t}^{(i,j)}$ into a denoised sub-CFR map $\tilde{\mathbf{M}}_{r,t}^{(i,j)}$. The full CFR map is reconstructed through 2D interpolation. It should be emphasized that the denoising process is executed each time an $M_f \times M_t$ noisy sub-CFR map becomes available during inference.

    \begin{algorithm}[t]
        \caption{Proposed Online Adaptive Channel Denoising Procedure}\label{alg:Procedure}
        \small
        \begin{algorithmic}[1]
            \State \textit{Fine-Tuning Phase:}
            \FOR{$s=1$ to $N_{\rm train}$}
                \State Perform conventional DM-RS channel estimation in slot $s$.
                \State Perform data detection in slot $s$. 
            \ENDFOR
            \State Determine the optimal size $(P^\star,Q^\star)$ from \eqref{eq:MSE}.
            \State Compute data-aided sub-CFR estimates for $s \in \{1, \ldots, N_{\rm train}\}$ from \eqref{eq:DA_CE}.
            \State Construct the online training dataset $\mathcal{D}_{\rm online}$ from \eqref{eq:onlineset_DL}.
            \State Fine-tune the pre-trained DnNN using $\mathcal{D}_{\rm online}$.
            \State \textit{Inference Phase:}
            \FOR{$s=N_{\rm train}+1$ to $N_{\rm total}$}
                \State Perform conventional DM-RS channel estimation in slot $s$.
                \State Extract noisy sub-CFR estimates via sub-sampling. 
                \IF {Collect $M_{t}$ consecutive sub-CFRs in the time domain}
                    \State Construct the noisy sub-CFR map of size $M_{f} \times M_t$.
                    \State Denoise the sub-CFR estimates using the fine-tuned DnNN.
                    \State Reconstruct a full CFR map via 2D interpolation.
                    \State Perform data detection using the reconstructed CFRs.
                \ENDIF
            \ENDFOR          
        \end{algorithmic}
    \end{algorithm}

    \section{Simulation Results}\label{Sec:simulation}
    In this section, we evaluate the superiority of the proposed channel denoising methods through Monte Carlo simulations by measuring the frame error ratio (FER) and the normalized mean squared error (NMSE) gain. 
    
    \subsection{Simulation Setup}
    In our simulations, we consider a MIMO-OFDM system operating at a center frequency of $\rm{3.5GHz}$ with a subcarrier spacing of $\rm{15kHz}$ and a total of $512$ subcarriers serving for $4$ users. Each slot consists of \(14\) OFDM symbols, i.e., $N_{\rm OFDM} = 14$, and each resource block  contains \(12\) consecutive subcarriers. We use 4-QAM for symbol modulation. The number of transmit and receive antennas is set to $(N_t, N_r) = (2, 16)$. For channel coding, we adopt a low-density parity-check code with a code rate of $1/2$.
    The SNR of the system is defined as $\mathsf{SNR} = \frac{N_t}{\sigma^2}$, and the NMSE is defined as 
    \begin{align}
        {\rm NMSE}~ [{\rm dB}] = 10\log_{10} \frac{ \mathbb{E} \left[ \|\hat{\bf H}[n,k] - {\bf H}[n,k] \|_{\mathrm F}^2\right]}{\mathbb{E} \left[  \|{\bf H}[n,k]\|_{\mathrm F}^2 \right]}.
    \end{align}

    For performance comparison, we consider the following methods:
    \begin{itemize}
        \item {\bf Perfect CSIR}:
            This is an ideal baseline in which perfect CFRs are assumed to be available at all REs.
            
        \item {\bf Conventional CE}:
            This is a conventional channel estimation method based solely on DM-RS, without channel denoising. The CFRs at DM-RS positions are estimated using the LS method, and a 2D linear interpolation is used to reconstruct the CFRs at non-DM-RS positions.
            
        \item {\bf Data-Aided CE}:
            This is a classical data-aided channel estimation method in \eqref{eq:DA_CE} in which detected data symbols are used as virtual DM-RSs, as discussed in Sec.~\ref{Sec: DA_CE}.
            
        \item \textbf{Transfer (Pre-Trained)}:
        This method uses a pre-trained DnNN trained on offline data. No fine-tuning is applied during online communication.
        
        \item \textbf{Meta (Pre-Trained)}:
        Similar to \textbf{Transfer (Pre-Trained)}, this method uses a DnNN trained via meta learning across multiple tasks. No fine-tuning is applied during online communication.

        \item \textbf{Transfer (Proposed)}:
        This is our online adaptive channel denoising method in Algorithm~\ref{alg:Procedure} when employing the proposed transfer learning approach. 
        
        \item \textbf{Meta (Proposed)}:
        This is our online adaptive channel denoising method in Algorithm~\ref{alg:Procedure} when employing the proposed  meta learning approach. 
    
        \item \textbf{Transfer (True CFR)}:
        This variant of \textbf{Transfer (Proposed)} uses true CFRs as labels for online training data.
    
        \item \textbf{Meta (True CFR)}:
        This variant of \textbf{Meta (Proposed)} also uses true CFRs as labels for online training data. 
    \end{itemize}
    
    For all the channel denoising methods considered, we apply a 2D linear interpolation technique to reconstruct the full CFR map from the denoised sub-CFR map.
    For both DM-RS-based channel estimation at DM-RS positions and data-aided channel estimation, we use the LS method, while the LMMSE method is adopted for data detection.

    Among various DnNN architectures, we adopt the denoising convolutional neural network proposed in~\cite{dncnn}, which is well known for its effectiveness in image denoising tasks. This architecture is trained to learn a residual mapping $\mathcal{R}(\hat{\bf M}_{r,t}^{(i,j)}; \theta_{\rm DL})$ that estimates the noise component in the input. To this end, we define the loss function following the formulation in~\cite{dncnn}:
    \begin{align}
        \ell(\hat{\bf M}_u,{\bf M}_u;\theta) = \big\|\mathcal{R}(\hat{\bf M}_{u}; \theta) - (\hat{\bf M}_{u} - {\bf M}_{u})\big\|_{\mathrm F}^2.
    \end{align}
    The residual learning formulation encourages the neural network to directly predict the noise, which is then subtracted from the input to recover the denoised channel. We set the depth of the denoising convolutional neural network to $5$ layers.

    We set the total number of slots to $N_{\rm total} = 8$, of which the first $N_{\rm train} = 4$ slots are used for online adaptation (i.e., fine-tuning) and the remaining slots are used for performance evaluation during inference. In addition, we fix the window size for the denoising model input as $(M_t, M_f) = (8, 58)$ corresponding to the number of OFDM symbols and subcarriers included in each sub-CFR map.
    For the optimization of the window size parameters $P$ and $Q$, we adopt the criterion defined in \eqref{eq:MSE}, assuming uniform reliability weights $r_{n,k} = 1$ for all $(n,k)$. We define a set of candidate integer values for $P\in\{ 1, 2, 4, 7\}$ and $Q \in \{ 1, 2, 3, 6\}$, and select the optimal combination that minimizes the objective.

    We construct an offline dataset $\mathcal{D}_{\rm off} = \{(\hat{\bf M}_u, {\bf M}_u)\}_{u=1}^{N_{\rm off}}$ consisting of $N_{\rm off} = 30{,}000$ training samples. Among these, $27{,}000$ samples are used for training and $3{,}000$ samples are used for validation. To enhance robustness, the samples are generated under multiple SNR conditions, specifically $\mathsf{SNR} \in \{-6, -3, 0, 6, 9\}$ dB. For both training and testing, we consider the CDL channel model defined in the 3GPP specification \cite{cdl}. The CDL model represents multipath propagation by combining multiple clusters, each of which consists of several rays with similar delay spreads. Due to its realistic spatial-temporal structure, the CDL model is widely adopted for link-level performance evaluation in wireless communication systems.

    \begin{itemize}
        \item {\bf Transfer Learning (Pre-training):} For the transfer learning setting, the DnNN is pre-trained using the offline dataset described above. Training is carried out over $E_{\rm T} = 100$ epochs using the ADAM optimizer, with a batch size of $B_{\rm T} = 64$ and a learning rate of $10^{-3}$. This pre-training phase allows the DnNN to learn general denoising features from a wide range of channel conditions, enabling it to provide a strong initialization for subsequent online fine-tuning. 

        \item {\bf Meta Learning (Meta-training):} For the meta learning setting, we adopt an episodic training strategy based on the MAML algorithm. Each task is constructed with $N_{\rm sup} = 16$ support samples and $N_{\rm que} = 32$ query samples. Given a $N_{\rm off} = 30{,}000$ offline samples, we construct $625$ meta-training tasks. The model is meta-trained for $E_{\rm M} = 100$ epochs, with a task batch size of $B_{\rm M} = 16$. We set the inner-loop learning rate to $\alpha = 10^{-3}$ and the outer-loop learning rate to $\beta = 10^{-3}$. Through this procedure, the model learns a globally adaptable initialization that can be rapidly fine-tuned to new channel environments using only a small number of online samples. 
    \end{itemize}

    In these simulations, we consider two scenarios: 
        \begin{itemize}
        \item {\bf In-distribution (ID) scenario:} In this scenario, both training and testing are conducted using the CDL-B channel model with a $300{\rm ns}$ delay spread and a velocity of $120{\rm km/h}$. This scenario allows us to assess the baseline performance of each method without distribution shift. 

        \item {\bf Out-of-distribution (OOD) scenario:} In this scenario, both the DnNN parameters are pre-trained on the CDL-E channels with a delay spread of $300\mathrm{ns}$ and the velocity of $120\mathrm{km/h}$, but tested on the CDL-B channel with a delay spread of $300\mathrm{ns}$ and the velocity of $120\mathrm{km/h}$. This scenarios allows us to test mismatched channel environment.
    \end{itemize}

    \subsection{Performance Evaluation}
    \begin{figure}[t]
        \centering
        \includegraphics[width=8cm]{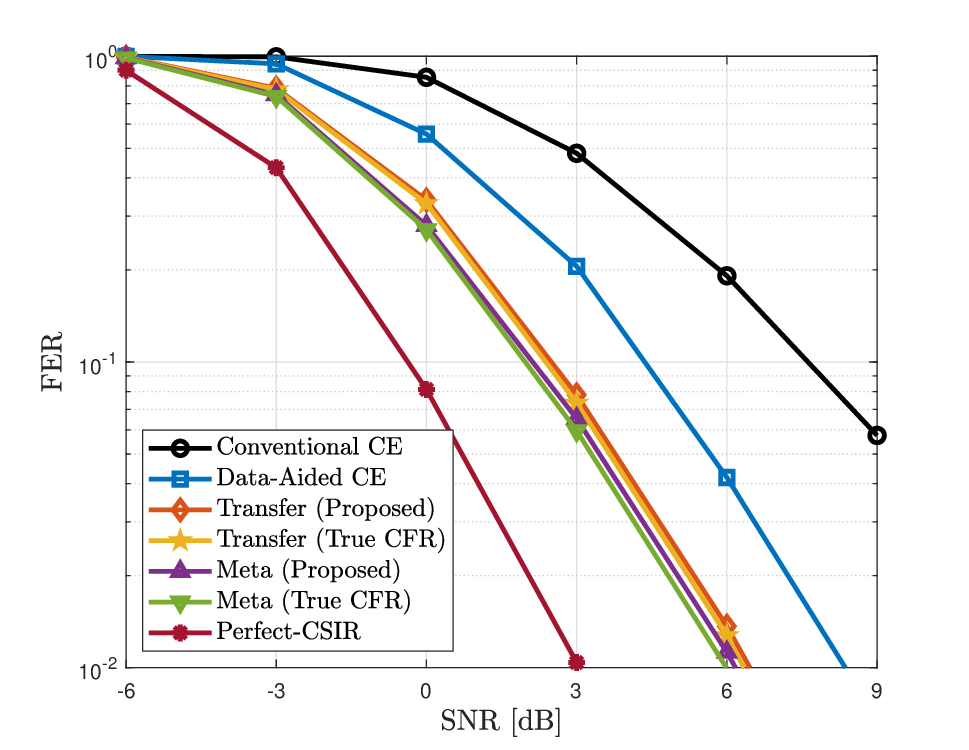}
        \caption{FER comparison of various channel estimation and denoising methods under the ID scenario.} 
        \vspace{-3mm}
        \label{fig: FER_ID}
    \end{figure}
    
    Fig. \ref{fig: FER_ID} compares the FERs of various channel estimation and proposed denoising methods under the ID scenario. During the fine-tuning stage, the transfer learning model is fine-tuned using $30$ gradient updates with a learning rate of $10^{-3}$, while the meta learning model uses only $10$ updates with a learning rate of $10^{-4}$. In both cases, $32$ online samples are used for fine-tuning. 
    As observed in Fig. \ref{fig: FER_ID}, the proposed methods, Transfer (Proposed) and Meta (Proposed), consistently outperform the conventional baselines, Conventional CE and Data-Aided CE, across all SNR levels. Furthermore, the performances of Transfer (Proposed) and Meta (Proposed) closely approach those of Transfer (True CFR) and Meta (True CFR), respectively. These results confirm the effectiveness of the proposed online data generation strategy, even in the absence of ground-truth labels.

     \begin{table*}[t]
        \centering
        \caption{FER comparison across different $(P, Q)$ and $\mathsf{SNR}$ levels under the OOD scenario.}
    
        \begin{subtable}[t]{0.48\textwidth}
            \centering
            \caption{Transfer Learning}
            \begin{tabular}{c|ccccc}
                \toprule
                $(P, Q) \backslash \mathsf{SNR}$  & $-3$ & $0$ & $3$ & $6$ & $9$ \\
                \midrule
                $(1,1)$   & 0.8667 & 0.4003 & 0.0898  & 0.0146 & 0.0015 \\
                $(2,1)$   & 0.8378 & 0.3656 & 0.0765  & 0.0120 & 0.0014 \\
                $(2,2)$   & 0.8260 & 0.3536 & 0.0747  & 0.0121 & \textbf{0.0012} \\
                $(2,3)$   & 0.8069 & \textbf{0.3346} & \textbf{0.0701}  & \textbf{0.0108} & 0.0012 \\
                $(4,3)$    & 0.7992 & 0.3433 & 0.0776  & 0.0149 & 0.0034 \\
                $(7,6)$    & \textbf{0.7827} & 0.3417 & 0.0772  & 0.0144 & 0.0038 \\
                \bottomrule
            \end{tabular}
            \label{tab:fer-transfer}
        \end{subtable}
        \hfill
        \begin{subtable}[t]{0.48\textwidth}
            \centering
            \caption{Meta Learning}
            \begin{tabular}{c|ccccc}
                \toprule
                $(P, Q) \backslash \mathsf{SNR}$ & $-3$ & $0$ & $3$ & $6$ & $9$ \\
                \midrule
                $(1,1)$    & 0.8390 & 0.3949 & 0.1085 & 0.0230 & 0.0073 \\
                $(2,1)$   & 0.8203 & 0.3677 & 0.0964  & 0.0204 & 0.0070 \\
                $(2,2)$    & 0.8100 & 0.3582 & 0.0933  & 0.0195 & \textbf{0.0067} \\
                $(2,3)$    & 0.8001 & 0.3448 & \textbf{0.0878}  & \textbf{0.0191} & 0.0068 \\
                $(4,3)$    & 0.8016 & 0.3503 & 0.0935  & 0.0215 & 0.0077 \\
                $(7,6)$    & \textbf{0.7869} & \textbf{0.3333} & 0.0894  & 0.0202 & 0.0073 \\
                \bottomrule
            \end{tabular}
            \label{tab:fer-meta}
        \end{subtable}
    
        \label{tab:fer-subtables}
        \vspace{-3mm}
    \end{table*}   
    \begin{figure}[t]
        \centering
        \includegraphics[width=8cm]{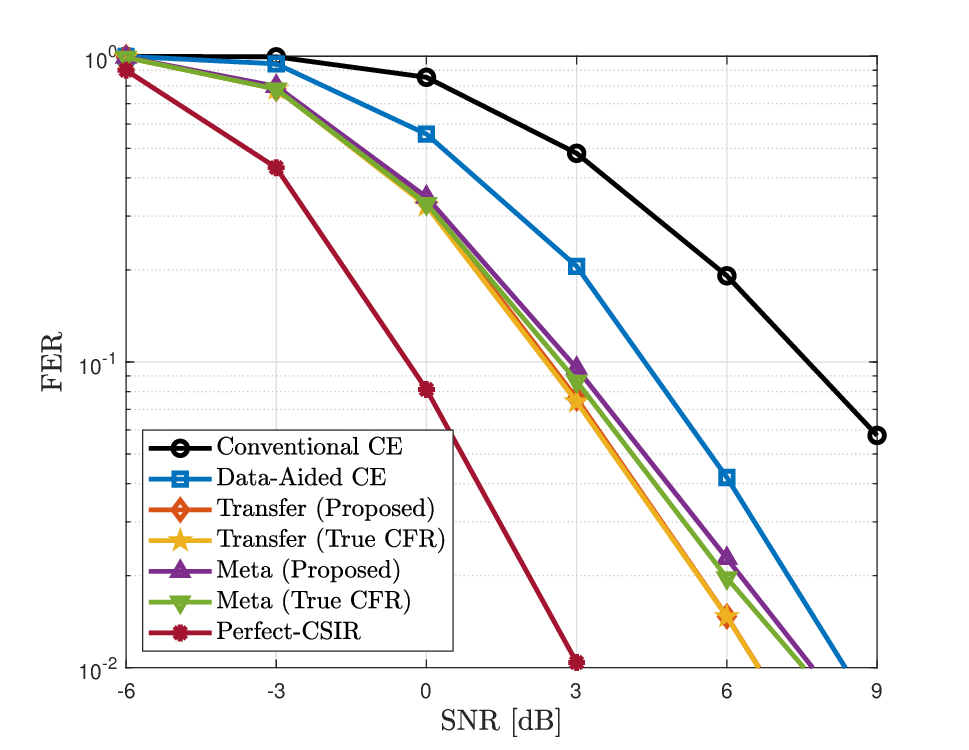}
        \caption{FER comparison of various channel estimation and denoising methods under the OOD scenario.} 
        \vspace{-3mm}
        \label{fig: FER_OOD}
    \end{figure}

    
    Fig. \ref{fig: FER_OOD} compares the FERs of various channel estimation and denoising methods under the OOD scenario. The fine-tuning configurations are identical to those used in Fig. \ref{fig: FER_ID}.
    Despite the change in channel characteristics, both Transfer (Proposed) and Meta (Proposed) demonstrate strong robustness, achieving significantly lower FER compared to the Conventional CE and Data-Aided CE baselines. Moreover, the performances of both methods closely approach those of their ideal variants, Transfer (True CFR) and Meta (True CFR), as observed in Fig. \ref{fig: FER_ID}.
    Notably, the meta learning-based method achieves this performance with fewer gradient updates, highlighting its ability to quickly adapt to previously unseen channel conditions. This highlights the strength of meta learning combined with the proposed online data generation strategy in enabling fast adaptation and generalization in dynamic wireless environments. Nonetheless, the transfer learning-based method can outperform meta learning when fine-tuned with a sufficient number of gradient steps and appropriately tuned hyper-parameters.
    
    
    \begin{figure}[t]
        \centering
        \includegraphics[width=8cm]{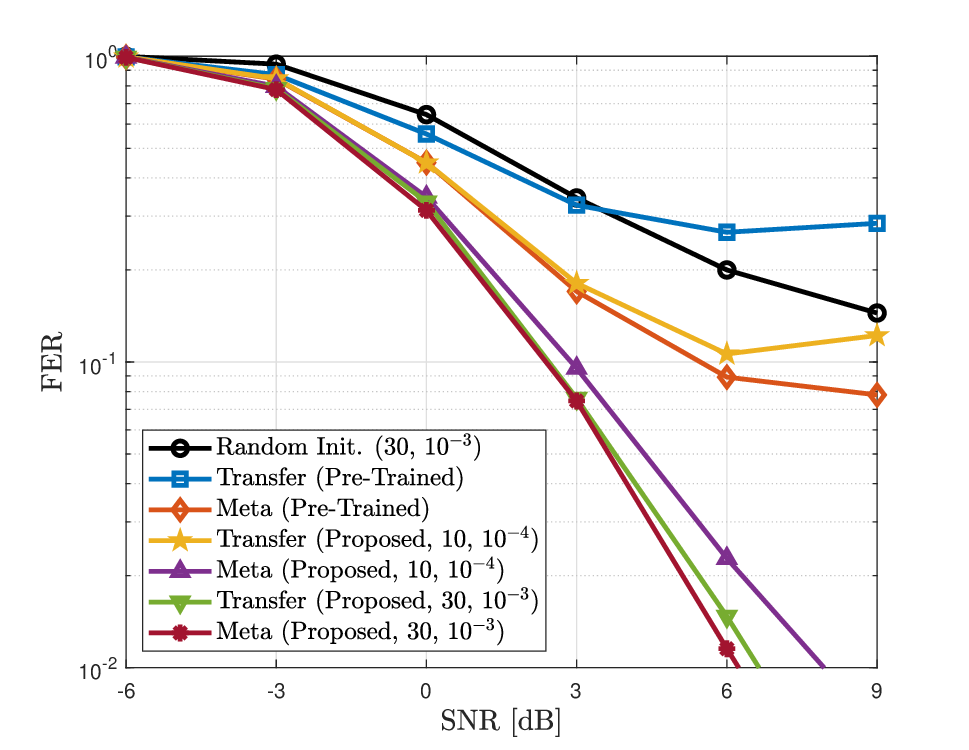}
        \caption{FER comparison of the proposed channel denoising methods with various fine-tuning strategies under the OOD scenario.}
        \vspace{-3mm}
        \label{fig: FER_finetune}
    \end{figure}
    
    Fig. \ref{fig: FER_finetune} compares the FERs of the proposed channel denoising models with various fine-tuning configurations under the OOD scenario. Specifically, we test the impact of (i) no fine-tuning, and (ii) different combinations of gradient update steps and learning rates during the online adaptation phase.
    First, we observe that {Transfer (Pre-trained)}, which uses a DnNN pre-trained on offline data without any fine-tuning, yields poor performance, especially at moderate-to-high SNR levels. This indicates that pre-training alone is insufficient for adapting to real-time channel variations under the OOD scenario. In contrast, {Meta (Pre-trained)} achieves better performance than {Transfer (Pre-Trained)}, demonstrating that the initialization learned via meta learning provides better generalization (i.e. good initial model parameter) even without further adaptation.
    When a small number of adaptation steps and low learning rate are used (i.e., 10 updates with learning rate $10^{-4}$), {Meta (Proposed, 10, $10^{-4}$)}, demonstrates noticeable improvement over {Meta (Pre-Trained)}, indicating effective adaptation even with limited updates resources. In contrast, {Transfer (Proposed, 10, $10^{-4}$)} fails to improve over its pre-trained version, suggesting that transfer learning is less robust under constrained fine-tuning settings.
    As the number of updates increases and a higher learning rate is employed (i.e., 30 updates with $10^{-3}$), both methods exhibit significant performance gains. In particular, {Transfer (Proposed, 30, $10^{-3}$)} reaches competitive FER performance. However, {Meta (Proposed, 30, $10^{-3}$)} still exhibits a better performance than the transfer learning approach, highlighting its strong adaptability with fewer fine-tuning efforts. 
    These results suggest that while transfer learning can eventually achieve strong performance with sufficient fine-tuning, the meta learning approach is more sample and gradient efficient, making it a practical choice for real-time adaptation scenarios with limited resources.

    Table \ref{tab:fer-subtables} compares the FERs of the proposed channel denoising methods under various combinations of $(P, Q)$ and SNR values. 
    As shown in the table, in low-SNR regimes (e.g., $\mathsf{SNR} = -3$dB), larger values of $(P, Q)$ tend to yield improved FER performance. This behavior suggests that incorporating more virtual DM-RSs helps compensate for severe noise levels, even if they are less correlated. Conversely, in high-SNR regimes (e.g., $\mathsf{SNR} = 6\sim 9$dB), smaller values of $(P, Q)$  become favorable, as incorporating only highly-correlated virtual DM-RSs are sufficient for accurate channel refinement.  Overall, the consistent performance trend observed in Fig.~\ref{fig: proposed receiver} and Table~\ref{tab:fer-subtables} substantiates the effectiveness of the proposed window size optimization strategy in Sec.~\ref{Sec:Opt_Range}. These findings confirm that careful selection of the time-frequency window size in the data-aided training process can significantly enhance channel estimation accuracy.
    
    \begin{figure}[t]
        \centering
        \includegraphics[width=8cm]{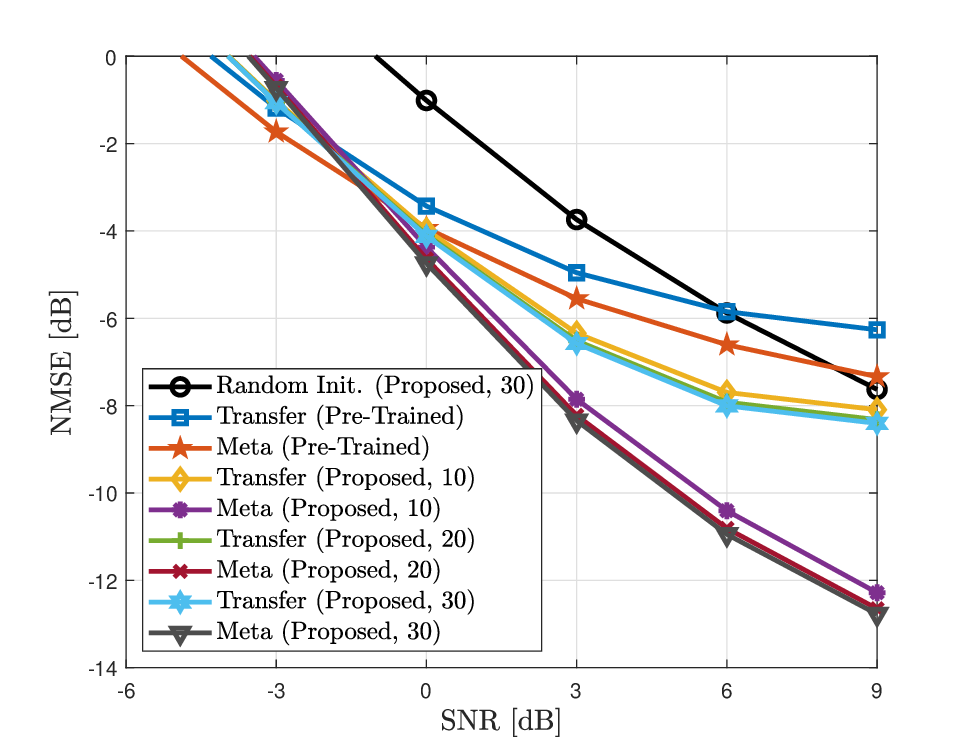}
        \caption{NMSE comparison of the proposed channel denoising methods with varying amounts of data for fine-tuning under the OOD scenario.}
        \vspace{-3mm}
        \label{fig: NMSE_sample}
    \end{figure}

    Fig. \ref{fig: NMSE_sample} compares the NMSEs of the proposed channel denoising models with varying numbers of online fine-tuning samples under the OOD scenario. The number in brackets indicates the number of online data samples used during fine-tuning. As shown in the figure, the proposed meta learning method achieves competitive performance even with as few as 10 samples, demonstrating its capacity to adapt efficiently when online training data is scarce. In contrast, the transfer learning method exhibits relatively limited performance improvement when only a small number of online samples are available. However, as the number of samples increases, the transfer learning method demonstrates a more pronounced performance enhancement compared to the meta learning method. These results highlight the distinct advantages of each method, demonstrating that the choice between transfer learning and meta learning should depend on both the number of available online training samples and the diversity of deployment scenarios.

    \begin{figure}[t]
        \centering
        \includegraphics[width=8cm]{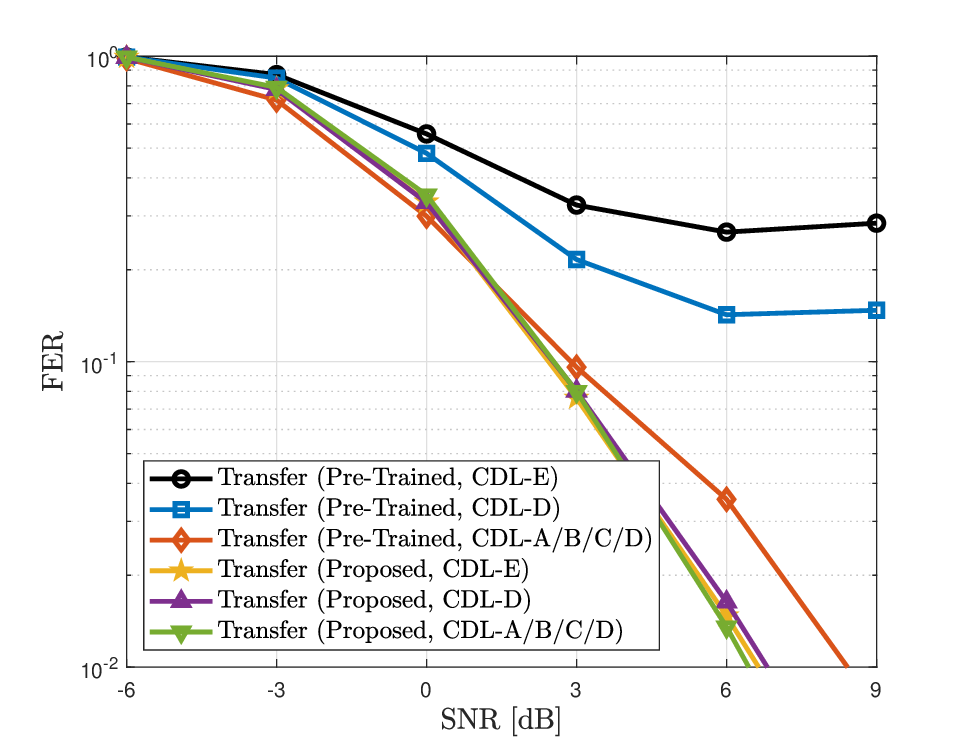}
        \caption{FER comparison under CDL-B for various pre-training.}
        \vspace{-3mm}
        \label{fig: FER_transfer}
    \end{figure}

    Fig. \ref{fig: FER_transfer} presents the FER of the proposed transfer learning method with different environments for pre-training. 
    The testing environment follows the CDL-B model specified in the OOD scenario. Transfer (Pre-Trained, CDL-D) and Transfer (Pre-Trained, CDL-E) exhibit relatively poor performance across all SNR levels, primarily due to the domain mismatch between the non-line-of-sight (NLOS) characteristics of CDL-D/E and the line-of-sight (LOS) nature of CDL-B.
    In contrast, Transfer (Pre-Trained, CDL-A/B/C/D), which is trained on a more diverse combination of LOS and NLOS channels, achieves noticeably better generalization to the CDL-B environment. 
    Notably, all the proposed methods demonstrate significant performance improvements over their respective pre-trained counterparts, especially in moderate-to-high SNR regimes. This confirms the effectiveness of the proposed online fine-tuning strategy in adapting to online channel environments.
    Although Transfer (Pre-Trained, CDL-A/B/C/D) already achieves competitive performance, its fine-tuned version further improves FER, particularly at high SNR levels. This result indicates that even well-generalized models can benefit from environment-specific adaptation when properly fine-tuned using the proposed strategy.

    \section{Conclusion}\label{Sec:Conclusion}
    In this work, we proposed an online adaptive channel denoising framework for MIMO-OFDM systems to overcome the limitations of conventional denoising methods that rely on offline training and thus fail to generalize across diverse channel conditions. We developed a standard-compatible online data generation strategy based on data-aided channel estimation, enabling supervised model adaptation without access to ground-truth channel information. Additionally, we provided an analytical formulation to optimize the time-frequency window size $(P, Q)$ employed in the data-aided estimation process. Utilizing the generated online dataset, we introduced two adaptive channel denoising approaches based on transfer learning and meta-learning techniques, respectively. Simulation results demonstrated that the proposed methods substantially outperform traditional baselines in terms of FER and NMSE under both ID and OOD scenarios. Furthermore, our findings revealed that the meta learning method offers robust adaptation capabilities with limited parameter updates, while the transfer learning approach achieves superior performance when sufficient online training data is available.
    
    Potential directions for future research include developing adaptive fine-tuning schedules that dynamically adjust learning rates and update steps according to real-time channel conditions such as SNR or mobility. In addition, exploring alternative denoising neural network may further improve performance.

\bibliographystyle{unsrt}

\end{document}